# Team performance and large-scale agile software development

Muhammad Ovais Ahmad[†]
Department of Computer Science
Karlstad University
Karlstad Sweden
ovais.ahmad@kau.se

Hadi Ghanbari
Department of Information and
Service Management
Aalto University
Espoo Finland
hadi.ghanbari@aalto.fi

Tomas Gustavsson
Karlstad Business School
Karlstad University
Karlstad Sweden
tomas.gustavsson@kau.se

## ABSTRACT

**Background**: Software development is a team work and largely dependent on open social interaction and continuous learning of individuals. **Objective**: Drawing on well-established theoretical concepts proposed by social psychology and organizational science disciplines, we develop a theoretical framework proposing that team climate (i.e. psychological safety, team cohesion, team structure, team relationships and interactions) has a significant influence on team learning and ultimately affects team performance. The goals of our study are 1) to understand the preconditions of team learning, and 2) to investigate the relationship between team learning, psychological safety, and team performance in large-scale agile software development projects. **Method:** We plan to conduct a survey with software professionals in Sweden's from three companies' partners in our large-scale agile research project.

## CCS CONCEPTS

• Software and its engineering • Agile software development.

## KEYWORDS

Agile Software Development, Large-Scale, Team Learning, Team Performance, Psychological Safety.

## 1 Introduction

In today's dynamic business environments, software companies strive to become more responsive in identifying and satisfying customers' needs. In doing so, the use of agile software development (ASD) has gained a tremendous amount of attention in the last two decades. Agile methodologies distinguish themselves from traditional development approaches by emphasising the importance of continuous value delivery and responsiveness to change [54], continuous communication and collaboration between stakeholders [1] and continuous reflection and learning [2, 3].

Continuous learning is a critical element in knowledge-intensive activities such as software development. Previous research shows that team learning has a significant influence on team performance [4-8], project success [2, 9], and a firm's innovation capabilities [10, 11]. In particular, the knowledge gained by individuals must be shared and integrated within teams to contribute to organisational learning [10]. As such, learning enables teams and organisations to reflect on their past decisions, and their outcomes, and continuously improve their practices and future decisions [3, 12].

Team learning is an interpersonal process that unfolds as software professionals and teams engage in knowledge sharing and reflection [2, 3, 11-13]. In other words, learning requires team members to engage in "asking questions, seeking feedback, experimenting, reflecting on results, and discussing errors or unexpected outcomes of actions'' [7](p. 353).

Reflection is one of the key aspects of the learning process [7] which is closely related to the team reflexivity construct [12], i.e. the ability of the team to (regularly) reflect and learn from their previous experiences [3]. This is crucial for software development teams to understand tasks, their relationships, actions and interactions. For instance, Scrum as an agile framework helps teams to discuss and share an understanding of their project activities and develop a shared mental model which results in improving performance [6]. Daily standup meetings and retrospectives assist a team in reflecting on how to be more efficient, after which it adjusts and adapts its behavior accordingly. In order to engage in continuous learning, team members need to feel safe.

Psychological safety is a necessary condition for team learning and performance [7]. Besides psychological safety, Dingsøyr et al [6] provided anecdotal support that team cohesion, learning, coordination, and goal-oriented leadership improve team performance.

According to Dikert et al. [14] large-scale ASD teams suffer from a range of challenges such as difficulties in implementing agile practices, integration of non-development teams, change resistance, coordination challenges in multi-team environment and so on [14]. When a team fail to collaborate, loses team bonding, and lacks continuous learning, the negative effects hinder agility in the medium and long term. Effective communication and coordination help in knowledge sharing and overall team learning [15, 16]. According to Bossche, Gijselaers, Segers, & Kirschner [17] team cohesiveness contribute to team learning. Other studies reported a positive correlation between team learning and team performance [18, 19].

While such factors have been extensively investigated in different contexts, we lack an understanding of the preconditions and mechanisms that influence learning in large-scale agile software development teams [6, 12, 20, 21]. The research is not clear on how exactly team learning and psychological safety are related to performance. Precisely, on whether the impact of team learning on performance is direct or mediated through other factors such as psychological safety, team cohesion, knowledge sharing, and team reflexivity. Especially there is a lack of empirical studies on preconditions of team learning in large scale ASD. As a first step in addressing this gap, we set out to investigate the preconditions of team learning in large-scale ASD projects by answering the following questions:

RQ1: What are the preconditions that influence team learning in large-scale ASD?

RQ2: What is the relationship between team learning, psychological safety, and team performance in large-scale ASD?

Drawing on well-established theoretical concepts proposed by social psychology and organisational science disciplines, we develop a theoretical framework proposing that team climate (i.e. psychological safety, team cohesion, team structure, team relationships and interactions) has a significant influence on team learning and ultimately affects team performance. To examine this framework, we will distribute an online survey among developers involved in large scale ASD. The data will be collected from three IT companies in Sweden.

## 2   Research Model and hypotheses development

Drawing on well-established theoretical concepts proposed by social psychology and organisational science disciplines, we develop a theoretical framework proposing that team climate (i.e. psychological safety, team cohesion, team structure, team relationships and interactions) has a significant influence on team learning and ultimately affects team performance. Team performance can be defined as the extent to which a team satisfies customer (or stakeholder) needs and expectations [7]. For a development team, several properties may be important, including adherence to predefined quality, and a schedule where certain deliverables are expected at predefined times. Considering our research question, there are ten hypotheses that we aim to test, described as follows.

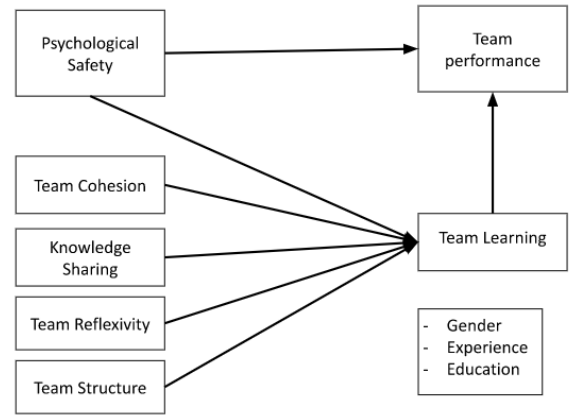

Figure 1: Proposed research model

**Psychological safety** is defined as "a shared belief held by members of a team that the team is safe for interpersonal risk-taking" [7] (p. 354). Previous research shows that psychological safety affects several team-level constructs including learning and performance. According to Edmondson [7], high levels of psychological safety have a positive influence on team performance. The author also shows that learning behavior mediates the relationship between team psychological safety and team performance. In another study, Carmeli at al. [22] shows that psychological safety is associated with higher levels of learning behaviors. We concentrate on team learning behaviors, which are routines by which a team acquires and processes knowledge, allowing it to flourish[7].

**Team learning behavior** is defined as gaining and sharing skills, knowledge, and information about work through the interaction of members [23]. These team member interactions involve such actions as experimentation, requesting feedback, and discussing errors [23]. To understand team learning, researchers highlighted examining both internal (i.e. capturing interactions among team members inside the team boundary) and external (i.e. capturing interactions beyond the team boundary with individuals and groups) learning behaviors [7, 23, 24]. Both internal and external learning have been empirically linked to team performance [4, 5, 7, 8]. This suggests thorough examination of psychological safety and team learning along with team performance in large-scale ASD context. Thus, we postulate that psychological safety enables learning and performance in large scale ASD projects:

**H1.** Psychological safety has a positive influence on team learning in large scale ASD projects.

**H2**. Psychological safety has a positive influence on team performance in large scale ASD projects.

**H3**. Team learning mediates the relationship between psychological safety and team performance in large scale ASD projects.

**Team Reflexivity** is "the extent to which group members overtly reflect upon the groups objectives, strategies and processes and adapt them to current or anticipated endogenous or environmental circumstances" [27]. Team reflexivity has a "dual focus" (cf. [28]), it involves both reflection upon previous accomplishments and adaptation to prepare for future actions. Schippers et al. [29] show that team reflexivity has an influence on team learning. Additionally, they concluded that team learning mediates the relationship between team reflexivity and performance. Higher team reflexivity predicts team effectiveness and team performance [30]. Based on these arguments we formulate the following hypothesis.

**H4**. Team reflexivity has a positive influence on team learning in large-scale ASD projects.

**H5**. Team learning mediates the relationship between team reflexivity and team performance in large-scale ASD projects.

**Knowledge sharing** refers to how "employees share their work-related experience, expertise, know-how, and contextual information with other colleagues" [31]. "Knowledge sharing as activities of transferring or disseminating knowledge from one person, group or organization to another" [32]. Knowledge sharing is a way to acquire cognitive resources (e.g., ideas, information, and knowledge) [33]. Working on a large-scale ASD project necessitates the transformation of internal and external knowledge into new forms. This can be achieved with continuous learning and knowledge sharing. According to Im and Rai [34], exploration and exploitation of knowledge sharing are both necessary and have an impact on performance. Knowledge sharing improves team performance in a range of ways such as improved decision making [35], better problem solving [36, 37] and enhanced creativity [38]. Team learning is also the outcome of communication and coordination, which allows team members to share knowledge about their team, work, resources, and context [15]. To develop knowledge and capacities and promote team performance, team members share knowledge, discuss, and reflect on mistakes [39, 40]. Team learning is positively related to knowledge sharing [16]. Hence, we propose that:

**H6**. Knowledge sharing has a positive influence on team learning in large scale ASD projects.

**H7**. Team learning mediates the relationship between knowledge sharing and team performance in large scale ASD projects.

**Team cohesiveness** is the degree to which team members like each other, identify themselves positively with the team, and want to remain its members [41]. Understanding group dynamics in teams requires cohesiveness [42]. Cohesive teams can foster knowledge sharing [43], enhances team learning [24], demonstrate increased collective efficacy [44], share responsibility for team failure [45], produce better software products than conflict-ridden teams [46], and achieve greater team success [47]. There is a positive relationship between team cohesion and team learning (e.g. [17]). Because learning behavior involves such actions as requesting feedback and discussing errors. This study approaches team learning as an outcome of cohesion that leads to improved team performance [7]. As a result, we propose that

**H8**. Team cohesion has a positive influence on team learning in large-scale ASD projects.

**H9**. Team learning mediates the relationship between team cohesion and team performance in large-scale ASD projects.

**Team structure** comes under the umbrella of team design and concerns specialization of tasks, hierarchical arrangements, and formalization of team objectives and procedures [48]. In any organization, the team structure varies because of need of the team, a particular context or specialization (48). The existing literature evidence that team structure play an important role in team learning [48], "organizational learning and innovation are promoted through less specialization, less hierarchy, and less formality, i.e., less structure" (e.g., [49]). Whereas Bunderson and Boumgarden [50] highlighted that team structure with clear role expectations, objectives, and priorities provides opportunities for team learning and promoting psychological safety. Team structure and empowering leadership contribute towards agility in software teams [51]. As a result, we propose that:

**H10**. Team structure has a positive influence on team learning in large-scale ASD projects.

## 3 Methodology

**Sample and Measures**

To test our theoretical model, we designed an online survey that will be distributed among software professionals working in three large Swedish software and systems development companies. We decided to follow a purposive sampling to recruit respondents who work in agile development teams in organizations that are involved in large-scale ASD projects. The companies use various Agile methods and practices in their development projects.

The questionnaire consists of three sections. In the first section, participants are provided information about the study and its objectives, and its expected benefits for the companies. The second section consists of 50 items adapted from previous literature for the seven research constructs. To assess Psychological Safety we used a 7-item scale developed by Edmondson [7]. Similarly, we adapted a 7-item Team Learning scale and a 5-item Team Performance scale from Edmondson [7]. To asses Team Cohesion we used the scale developed by Kakar [43]. We adapted a 3-item scale proposed by Kim & Lee [55] for Knowledge Sharing. For Team Reflexivity we used the 8-item scale proposed by Schippers et al., [29]. Finally, we adapted a 5-item Team Structure scale from Bresman & Zellmer-Bruhn [48].

Each item is evaluated using a seven-point Likert scale ranging from 1 (strongly disagree) to 7 (strongly agree). The final section of the questionnaire request demographic information from the participants. We will examine the moderating effect of gender, education, and work experience on the causal relationships shown in the model.

**Data analysis**

We will utilize Partial Least Squares Structural Equation Modeling to test our proposed model (PLS-SEM). It is noteworthy to highlight that PLS-SEM is the standard and well-established technique for investigating the link between variability and prediction models in quantitative research [52]. PLS-SEM provides much value for investigating causality based on prior theory [53], especially in behavioral research [56] similar to our study.

The PLS-SEM technique has two steps: the first includes evaluating the measurement model, and the second involves evaluating the structural model [52]. In the first step, we will assess the measurement model for internal consistency reliability, convergent validity, and discriminant validity [57, 58]. In the second step, the structural model will be assessed first by testing the hypotheses as direct effect. Next we will examine the direct, indirect, and mediating effects of the latent variables on Team Performance. Finally, we will conduct multi-group analysis to examine the influence of control variables on the results. To conduct the analysis we will use SPSS AMOS software.

**Threats to validity**

We use a self-reported survey to collect data from software professionals active in Swedish companies. Therefore, the results of the study may be influenced by response bias, since we cannot ensure that the respondents answer the questions truthfully and accurately. Additionally, even though the respondents and their companies have international backgrounds, the sample is not representative in terms of culture, projects and industrial domain.

## ACKNOWLEDGMENTS

This research is conducted under the Knowledge Foundation Sweden Grant and Karlstad University, Sweden.